# NVision-PA:
## A Tool for Visual Analysis of Command Behavior Based on Process Accounting Logs (with a Case Study in HPC Cluster Security)

Charis Ermopoulos[†‡]     William Yurcik[‡]

[†]Department of Computer Science
[‡]National Center for Supercomputing Applications  (NCSA)
University of Illinois at Urbana-Champaign (UIUC)
{*cermopo2,byurcik*}@ncsa.uiuc.edu

## Abstract

*In the UNIX/Linux environment the kernel can log every command process created by every user with process accounting. Thus process accounting logs have many potential uses, particularly the monitoring and forensic investigation of security events.  Previous work successfully leveraged the use of process accounting logs to identify a difficult to detect and damaging intrusion against high performance computing (HPC) clusters, masquerade attacks, where intruders masquerade as legitimate users with purloined authentication credentials.  While masqueraders on HPC clusters were found to be identifiable with a high accuracy (greater than 90%), this accuracy is still not high enough for HPC production environments where greater than 99% accuracy is needed.*

*This paper incrementally advances the goal of more accurately identifying masqueraders on HPC clusters by seeking to identify features within command sets that distinguish masqueraders. To accomplish this goal, we created NVision-PA, a software tool that produces text and graphic statistical summaries describing input processing accounting logs.  We report NVision-PA results describing two different process accounting logs; one from Internet usage and one from HPC cluster usage.  These results identify the distinguishing features of Internet users (as proxies for masqueraders) posing as clusters users. This research is both a promising next step toward creating a real-time masquerade detection sensor for production HPC clusters as well as providing another tool for system administrators to use for statistically monitoring and managing legitimate workloads (as indicated by command usage) in HPC environments.*

# 1.0  Introduction

Recent attacks enabled by stolen authentication passwords and unencrypted keys methods (sniffed via a key-logging console program, shoulder-surfed via bad security awareness, poor key management practices, etc.) have allowed intruders to masquerade as legitimate users on high performance computing (HPC) clusters.  This paper is the logical next step toward the development of a proactive alarm based on analysis of user command behavior in order to minimize damage from intrusions on high performance computing (HPC) clusters.  In a masquerade compromise, there is little to warn a security administrator that an account has been compromised since the attackers can properly authenticate so the intrusion may be long-lasting, persistent (difficult to reverse), and act as a stepping stone to more serious damage beyond a single user account. With the motivation of detecting masqueraders on HPC clusters, we have been working to discriminate different types of users based on their command behavior.

*Our intuition is that masqueraders act differently from legitimate HPC cluster users and the unique HPC cluster environment is constrained such that command behavior discrimination is enhanced versus enterprise environments.*  Since an HPC cluster environment should only have a minimal set of sanctioned system/application software available to users for performance purposes (software focused on supporting computation), the number of executable commands should be severely restricted.

Data from security operations reports validate this claim – while modes of attack on HPC systems vary greatly they do have one common characteristic, attacks are all very different from legitimate user activity found in a HPC cluster environment. For instance, once gaining access to a purloined account, attackers will typically communicate via an Internet Relay Chat (IRC), download additional exploits via ftp or the web, and then compile these exploits and attempt their execution. While ftp and compile commands may be typical in an HPC cluster environment, the command sequence pattern is the key.

In [9] we considered several methods to mitigate the threat from masqueraders and presented results from empirical testing showing that we can accurately discriminate enterprise users from cluster users based on their command behavior provided with a reasonable amount of training data (in terms of either number of commands or time period). Specifically in [9] we showed that by using Support Vector Machines (SVM) for classification with no constraints we are able to detect masqueraders with an accuracy of *94.9%*, together with a precision of *92.4%* and a recall of *91.9%*. Constraining the number of commands we found that as few as 10 commands provides an accuracy of *90%*. Constraining time we found that as short as 20 minutes had a precision over 75% with a recall slightly lower than *80%*. Examining the difference between the accuracy in number of commands versus time period reveals that the number of commands executed during time intervals varies significantly for different users. For instance, within 20 minutes, the user 'root' may execute hundreds of commands, but other users may only execute one command. Therefore where monitoring is measured by fixed time intervals, misidentification is generally higher due to a lack of observed commands for some users. The number of commands is the metric for classification while monitoring time is the window of observation.

While this previous work based on identifying masqueraders on HPC clusters using SVM classification provides good accuracy at *90%*, it is not good enough for a production HPC environment where the volume of users requires accuracy greater than *99%*. Although SVM classification trains on both individual commands and patterns of command usage, it does not provide necessary information on feature sensitivity which may be useful to increase masquerade detection accuracy.

In this paper we seek to identify the features of command behaviors useful for identifying masqueraders in HPC clusters environments so SVM classification techniques can be incrementally improved. The remainder of this paper is organized as follows: Section 2 presents the unique characteristics of process accounting as our data source. Section 3 provides an overview of the NVision-PA system architecture. Section 4 reports results from the use of NVision-PA on two processing accounting logs. We describe features for detecting masqueraders identified from NVision-PA statistical output. We end with conclusions and perspectives on future work in Section 5.

## 2.0  Background on Process Accounting

The UNIX/Linux accounting system collects information on individual/group usage of computer system resources. A system can record every process created by every user. This kind of logging is called *process accounting*. An example of the need for automated process accounting is the fact that many processes have short life spans that may escape human detection with *ps* command monitoring but still be of such high volume to dominate system load [1].

Process accounting has potentially high value for security purposes, for instance after a break-in to help determine what commands a user executed, correlating evidence, and incident investigation [3,5,7,9]. Other examples of security-related uses include:

- To hold a use accountable for some action indicated in the logs
- To enable the extraction of patterns of use of objects, users or security mechanisms in the system
- To identify security policy violations
- **To identify unsupported or vulnerable software is being used**
- To create an audit trail of the use (or abuse) that may occur from a specific user.
- To prevent the users from abusing the system by acting as a deterrent, given that the users know that there is a mechanism that logs security relevant actions in the system

While this work is motivated by the use of process accounting for security purposes, there are other uses for process accounting data. For example: (1) process accounting data is generally used in HPC environments to bill individual users (or groups of users) for the amount of CPU time that they consume [2,8] and (2) process accounting data provides an accurate source for workload characterization needed to tune applications and schedulers [1,4,6].

Process accounting is performed by the UNIX kernel. Every time a process terminates, the kernel writes a 32-byte record to the */var/adm/acct* or */var/adm/pacct* file that includes:

- name of the user and group who created the process
- first eight characters of the name of the command which launched the process
- elapsed time and processor time used by the process
- time that the process exited
- memory usage

• number of disk blocks read or written on behalf of the process
• flags, including:
    – S: Process was executed by the superuser
    – F: Process ran after a fork, but without an exec
    – D: Process generated a core file when it exited
    – X: Process was terminated by signal

The accounting file */var/adm/pacct* is accessed by many of the accounting utilities used with system accounting. For example the *lastcomm* program displays the contents of this file in a human-readable format. The *acctcom* utility is one of the most useful tools for getting a quick report from the system. It can be used to show all the processes that have been executed by a specific user, or to show all the processes, for any user, running longer than *x* seconds etc.

There are different process accounting software packages which introduce subtle variations in what we have generically described. The original was BSD Unix accounting, however, this does not have reliable messaging or consolidation reports. The open source Comprehensive System Accounting (CSA) package developed on Cray and IRIX platforms and now distributed by SGI consists of all accounting data for a given job identifier during a single system boot period. Red Hat Linux *psacct* accounting package contains several utilities for monitoring process activities including how long users have been logged on.

Process accounting data is subject to some inherent limitations with respect to security monitoring. For our purposes, we note two such limitations here. The first limitation is that process accounting does not keep track of parameters passed with the executed command. In fact, it only keeps track of the first eight characters of the command executed. Based on this, a malicious user could link a malicious tool to one with an innocent name and then execute the linked file.  The second limitation it is likely that an experienced attacker will attempt to delete any traces including command traces by suspending or stopping process accounting or modifying the existing process logfile of traces. To counteract the disabling of process accounting, it is possible to reliably and securely send the command history to protected servers at regular intervals where it can be archived beyond reach of intruders up to the point when the operating system is subverted and messages cease.

Lastly, there are two biasing effects with all process accounting software that must be considered: (1) "the Heisenberg Principle" – the processing needed to observe a system will impact the system, however, process accounting counters are always operational (whether turned on or off) with the processing impact occurring in post-processing when command history is flushed to the file system for analysis and (2) "the Edge Effect" – accounting records are written only for processes that have terminated so if a program runs for a long period it will not show up in the command history until the process is done [9]. While in this paper we focus on identifying features of  command behaviors useful for detecting masqueraders, in future work we plan to study both of these biasing effects: (1) determining the scalability limits of process accounting, both when does process accounting impact CPU performance and what activity level will impact process accounting performance and (2) determining how numerous and significant are "edge effect" processes.    For more details about process accounting, see [2,5,8,9].

## 3.0  NVision-PA System Architecture

NVision-PA is available from the following URL:

**<http://security.ncsa.uiuc.edu/distribution/NVision-PADownLoad.html>**

where there is also documentation on the installation procedures.

Figure 1 presents the system architecture of NVision-PA. The *Server* is the system from which we have collected the *Process Accounting Log File*, through the process accounting module that resides in the kernel of its operating system. The *NVision-PA Statistics Collector Engine* executes analysis processes on the log data as input and the results are sent to *NVision-PA GUI*, where they are organized into nine tabs and represented in text or graphics.

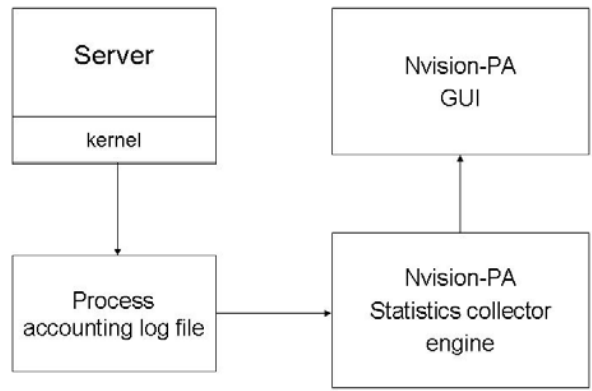

**Figure 1.** NVision-PA System Architecture

We have encountered two slightly different process accounting formats (and there are likely more we have not yet encountered). One format is found on Linux systems (*e.g. RedHat*) while the other one is found on UNIX systems (*e.g. Sun OS*). Our tool automatically recognizes either of these process accounting formats, parsing the data according to the recognized format. The Linux and UNIX formats differ mainly in the size of their respective fields (in bytes). Furthermore, the Linux format has additional fields (e.g. number of page faults) not found in the UNIX format. NVision-PA processes only the fields common to both formats.

## 3.1  NVision-PA Statistics Collector Engine

The Statistics Collection Engine is the core of NVision-PA. The collection engine parses the process accounting log file as input and produces a number of statistical reports for display through the NVision-PA GUI. When designing this engine, we had to consider two main factors: scalability and extensibility. Scalability is the primary requirement since the size of those logs can grow toward Terabytes depending on the volume usage of the system logging and the period of observation. To satisfy scalability, we generate nine different reports after only one pass through the log file. A secondary requirement is extensibility to add new reports or different statistics should new fields be added or new statistics become desirable – minimal effort should be required in order to produce a new report.

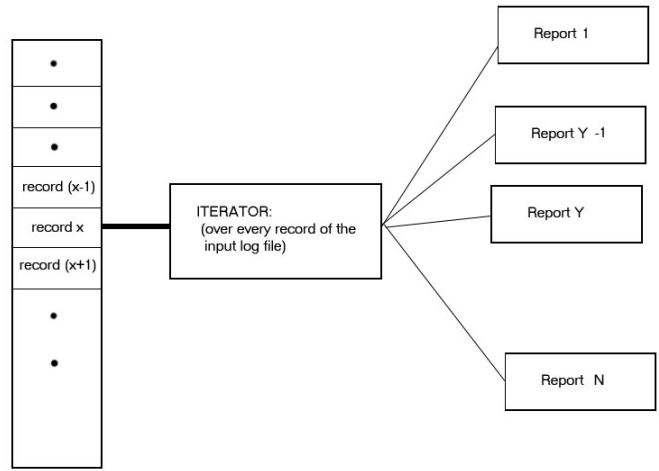

**Figure 2.** Statistics Collector Engine (Phase A)

Figure 2 shows the first part of the Statistics Collector Engine or Phase A.   In this phase an *iterator* object, reads the input log file, parses each individual process accounting record from the binary format that it is found in the file and sends each record for processing to each of the nine different *registered* reports.   For extensibility, we only need to parse the process accounting log file once regardless of the number of reports that we wish to generate as long as each different report is *registered* with the *iterator* in an initialization phase.

Figure 3 depicts when the *iterator* has finished reading the entire process accounting log file, each different *registered* report generates output sent it to a Statistics Collector. The Statistics Collector acts as an interface between the reports that generate raw output data and the GUI that renders the output displayed on the screen. We call this second part of the Statistics Collection Engine, Phase B.

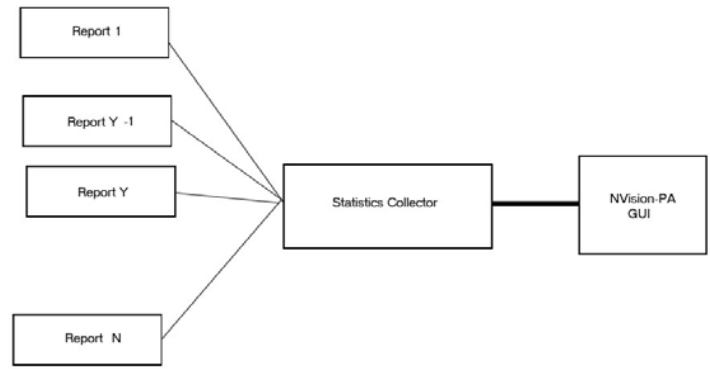

**Figure 3.** Statistics Collector Engine (Phase B)

The decision was made to produce NVision-PA in Java since process accounting logs can be found on many different platforms and we would like this tool to be portable for all of those platforms. For graphics we used the open source library *"Chart2D"* (a java library for drawing two dimensional charts by Jason J. Simas) that is available under the GNU license.[1]

## 3.2 NVision-PA Input

Figure 4 is a screenshot of the NVision-PA GUI showing an overlapping initial browse window through which the user can select by browsing or filename the exact process accounting log file to be selected as input. Format check occurs upon this input file selection. After a file is selected and format verified, processing for all reports is executed.

---

[1] <http://chart2d.sourceforge.net/>

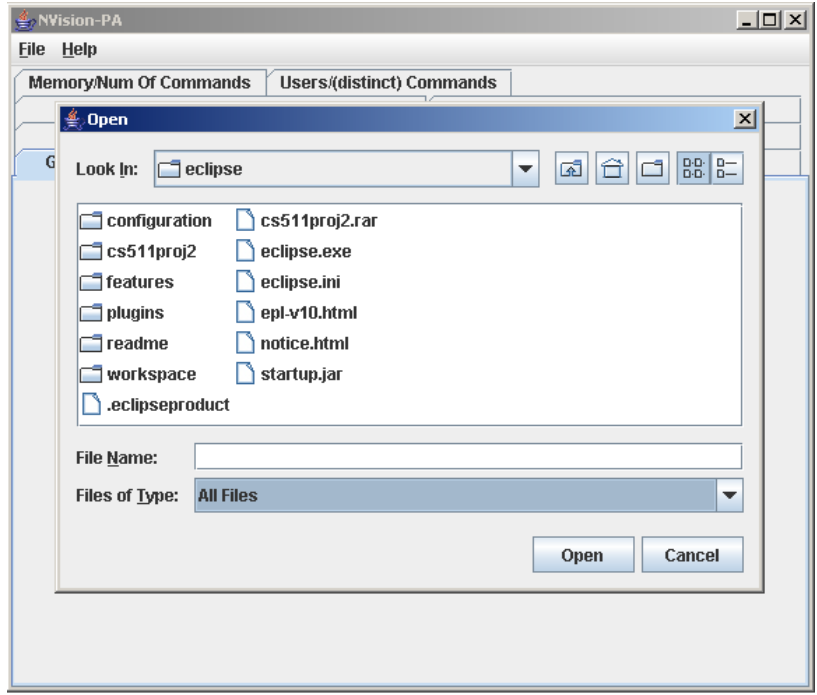

**Figure 4**  NVision-PA GUI for Input Selection

## 4.0 NVision-PA Output Results

There are nine different output reports implemented in the current version NVision-PA, each is represented as a tab on the NVision-PA GUI. In this section we display and discuss results from NVision-PA for each of these nine different output reports for the case of process accounting logs from two different sources: (1) a multi-user Internet server and (2) a multi-user HPC cluster.  Specifically we wish to compare how Internet server command behavior output differs from HPC cluster command behavior output for each of the nine reports. We also discuss whether these process accounting observations may be generalizable beyond these specific input data sets.

In Figure 5 we see an example of the NVision-PA "General" tab results which contains high-level information about the total number of commands, total number of distinct commands, the starting and ending date of the log, and effective logging period measured in days. The desired insight of this tab report is a general overview of the selected input processing accounting log file. In Figures 5A and 5B we see that both the Internet server and the HPC cluster processing accounting log files refer to about a month of data (31 and 28 days respectively) which is approximately equalized for comparison. We can conclude that the HPC cluster has more user interactivity over this period since it has 21 times more executed commands (1,853,411 versus 87,137). We can also conclude that the Internet server has a wider variety of the executed commands proportional to its usage since the ratio of distinct commands executed to total number of commands executed is an order of magnitude higher for the Internet server (.002) than the HPC cluster (.0004). "Distinct commands" in this context are defined as commands that occur at least once in the log file, multiple instances of the same command are not counted – a command is either present in the log file and counted once or not present/not counted. "Total commands" in this context are defined as each command execution instance, the same command executed $x$ times is counted $x$ times.

Figure 6 presents the NVision-PA "Users (non-distinct) Commands" tab results that refer to the distribution of total commands over all users in the selected input log. The desired insight from this tab report is the distribution of total commands[2] per user executed on the system during the period contained in the log. From this we can infer different types of

---

[2] Each command execution instance is counted, the same command executed $x$ times is counted $x$ times.

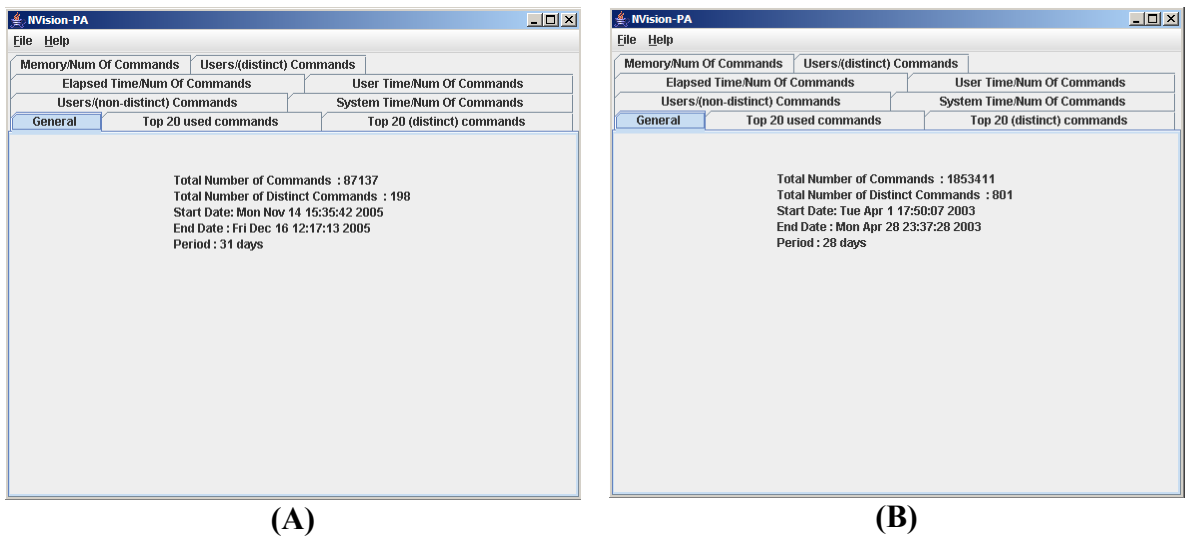

**(A)**                                **(B)**

**Figure 5.** NVision-PA General Information About PA Log File Including Total Non-Distinct Commands, Total Distinct Commands, and Time Period: (A) Internet PA Log File versus (B) HPC Cluster PA Log File.

users based on their command behavior. Figure 6A shows the distribution of users on the Internet server are clustered either at low command usage (about 40 users executed 0-20/20-40 commands within the log) or high usage (about 35 users executed 150-500/>500 commands within the log). Figure 6B shows the distribution of users on the HPC cluster are bimodally clustered at the extremes; low usage (about 20 users executed between 20-40 total commands) and high usage (about 70 users executed between 150-500/>500 total commands).

Figure 7 presents the NVision-PA "Users (distinct) Commands" tab results that refer to the distribution of distinct commands over all users in the selected input log. The desired insight from this tab report is the distribution of distinct commands[3] per user executed on the system during the period contained in the log.  From this we can infer different types of users based on the range of different commands executed – a user executing a small number of distinct commands is likely to be less capable than a user executing a wide variety of distinct commands.   Figure 7A shows the distribution of users on the Internet server is clustered at the medium range of distinct command usage (about 88 users executed between 10-15/15-20/20-25 distinct commands).   Figure 7B shows the distribution of users on the HPC cluster are bimodally clustered at the medium range of distinct command usage (about 25 users executed between 15-20 distinct commands) and at the high range of distinct command usage (about 50 users executed > 30 distinct commands). From this we infer that the average HPC cluster users are more capable in different command usage in the sense that they typically execute more distinct commands than the Internet server users.

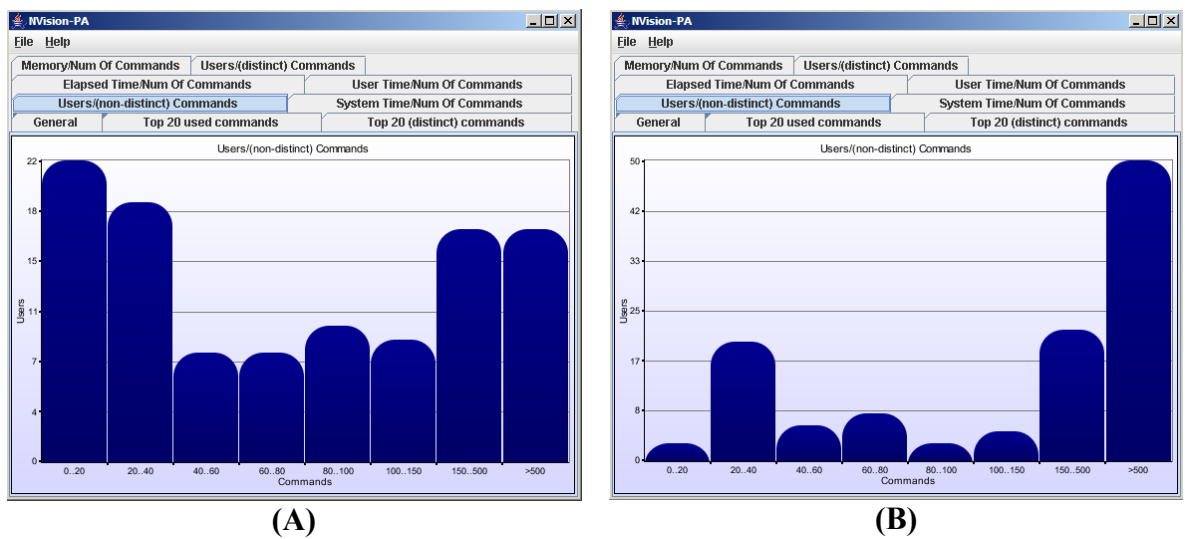

**Figure 6.** NVision-PA Statistical Distribution of Non-Distinct Commands: (A) Internet PA Log File versus (B) HPC Cluster PA Log File.

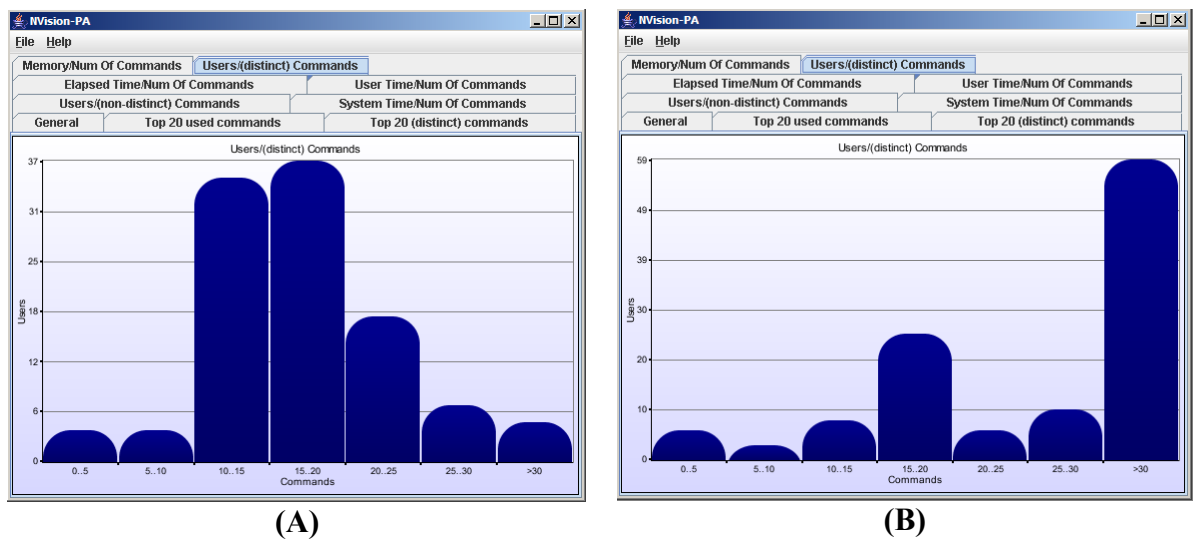

**Figure 7.** NVision-PA Statistical Distribution of Distinct Commands: (A) Internet PA Log File versus (B) HPC Cluster PA Log File.

Figure 8 presents the NVision-PA "Top 20 Used Commands" tab results that identify the most frequent instances of command execution over all users in the selected input log[4]. The desired insight from this tab report is the most commonly used commands in different environments.   The top 20 commands on the Internet server is lead by *sshd* remote login, then self-identification commands (*uname, hostname*) and also mail commands (*mail, elm, pine*), neither of these command types would be expected to occur in a HPC cluster environment.  The top 20 commands on the HPC cluster are dominated by shell commands (also commonly used on Internet servers) but also includes *pbs* referring to a specialized cluster scheduler, *rsync* referring to a specialized file transfer application, *sleep* referring to delay for a specified amount of time, and *libtool* for using shared libraries. Masquerader commands would not necessarily show up in the HPC cluster top 20 list since their frequency would be too low, however, automated Internet server processes (such as a mail spammer process) would be easily discernable as suspicious if they did appear.

---

[3] Multiple instances of the same command are not counted – a command is either present in the log file and counted once or not present/not counted, e.g. the same command executed *x* times is counted *only once*.
[4]  Each command execution instance is counted, the same command executed *x* times is counted *x* times.

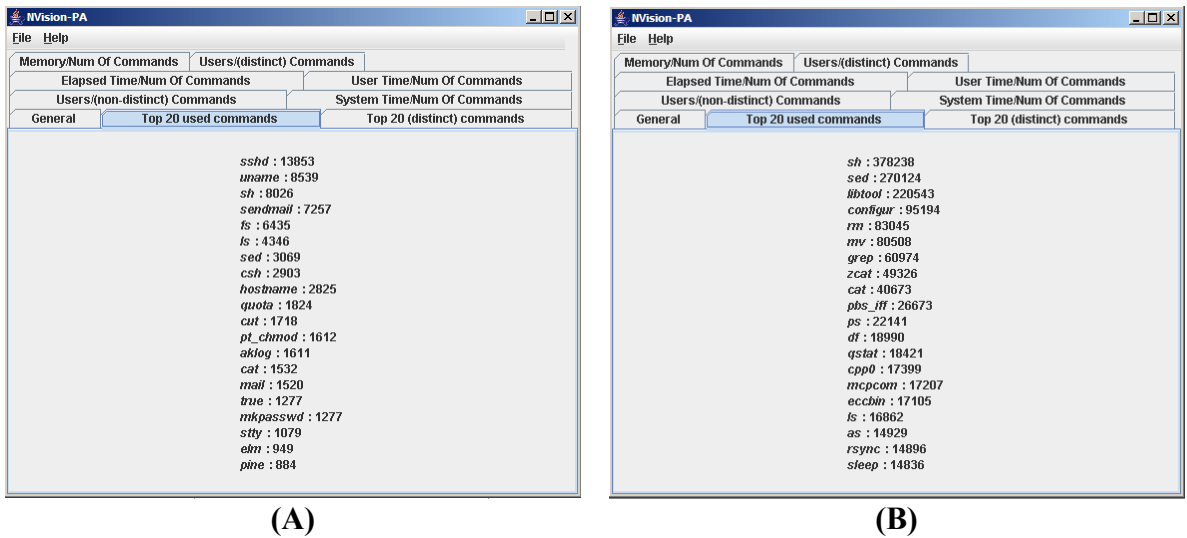

**Figure 8.** NVision-PA Top 20 Most Frequently Used (Non-Distinct) Commands: (A) Internet PA Log File versus (B) HPC Cluster PA Log File.

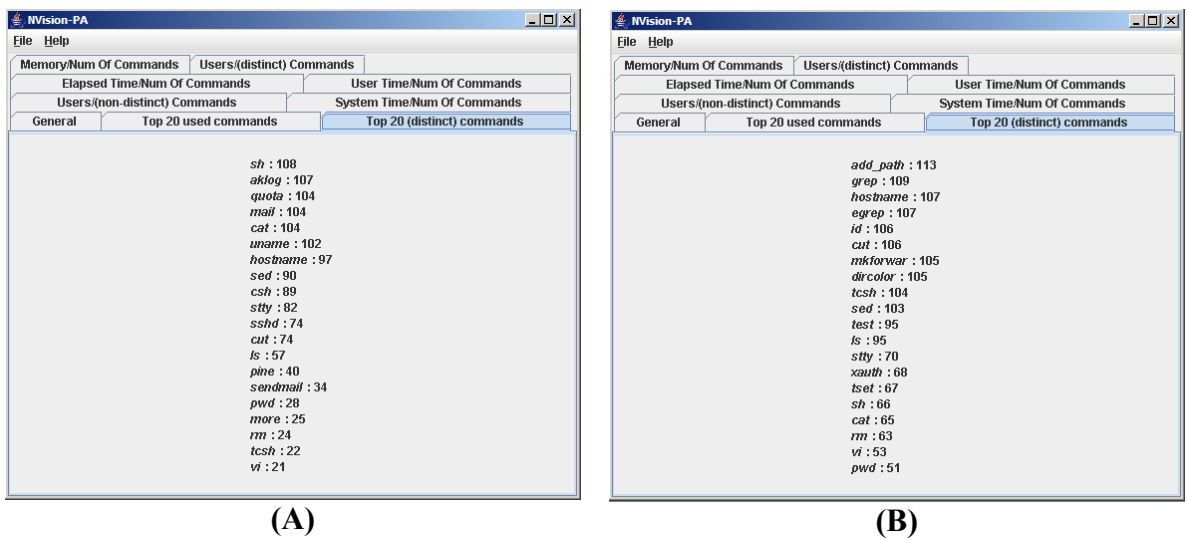

**Figure 9.** NVision-PA Top 20 Most Frequently Used (Distinct) Commands: (A) Internet PA Log File versus (B) HPC Cluster PA Log File.

Figure 9 presents the NVision-PA "Top 20 (distinct) Commands" tab results that identify the most frequently used distinct commands across different users.[5] The desired insight from this tab report is which commands have the widest user base. The top 20 distinct commands on the Internet server are again predominantly shell commands but also include *sshd* remote login, self-identification commands (*uname, hostname*), mail commands (*mail, pine, sendmail*), and editor commands (*vi*). The top 20 distinct commands on the HPC cluster are again predominantly shell commands but one sticks out – *68* HPC cluster users executed the *xauth command* (used to extract authorization records from one machine and merge with another for remote logins or granting access to other users). *xauth* would not be an advisable command to use outside of a trusted environment so it would be unlikely to appear in the Internet server command lists. Both "top 20" tab results also identify commands that are executed through ".profile" files that are potentially executed at every login of every user. For example, in Figure 9a we see *sh* and *aklog* and Figure 9b we see *addpath* which are likely executed as part of .profile.

---

[5] Regardless of how many instances an individual user executed a specific command, the total frequency of a distinct command will increase only by one if an individual user has ever executed that command.

Figures 10-12 refer to distribution of time parameters over commands. The time parameter is divided into system time, user time, elapsed time, and defined within process accounting with the following parameters:

> *ac_stime: system time spent in kernel space for the process (accurate to 0.01 second)*
> *ac_utime: user time spent in user space for the process*
> *ac_etime: total elapsed time for the process (greater than or equal to ac_utime + ac_stime)*
> *also known as "wall clock" or real-time (accurate to a second)*
> *{Note: these fields are reported in seconds with different accuracy}*

The relationship between these time components is as follows:

> *System Time(ac_stime) + User Time(ac_utime) <= Elapsed Time(ac_etime)*

Figure 10 presents the NVision-PA "System Time/Number of Commands" tab results that refer to the distribution of system time over all commands[6] in the selected input log. In other words, we can see how many commands had a system time of $x$ seconds. The desired insight from this tab report is the distribution of system time per command for different environments during the period contained in the selected input log. We may expect computational science researchers in the HPC cluster environment to have processes with higher system times since higher system times may reflect simulations and other tasks that the average Internet server user would not likely execute. The percentage of commands that executed with a system time of less than 0.1 seconds is comparable between the Internet server and HPC cluster environments; 14% and 9% respectively. The peak for both environments is commands executing with a system time between 1-2 seconds; 36% of the commands in the Internet server environment and 19% of the commands in the HPC cluster environment – roughly double the number of Internet server commands executed with a system time of between 1-2 seconds than HPC cluster

---

[6] Each command execution instance is counted, the same command executed $x$ times is counted $x$ times.

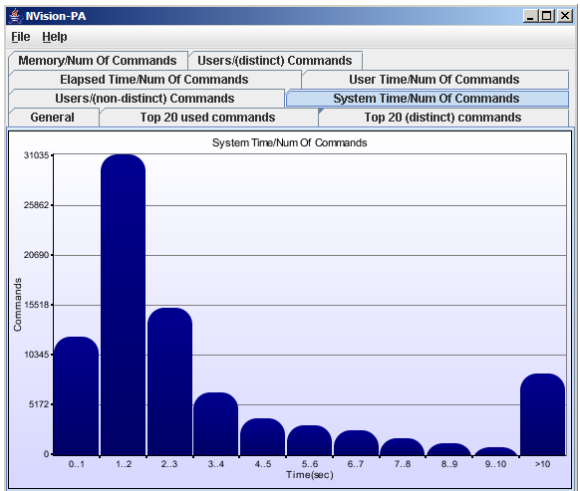
**(A)**

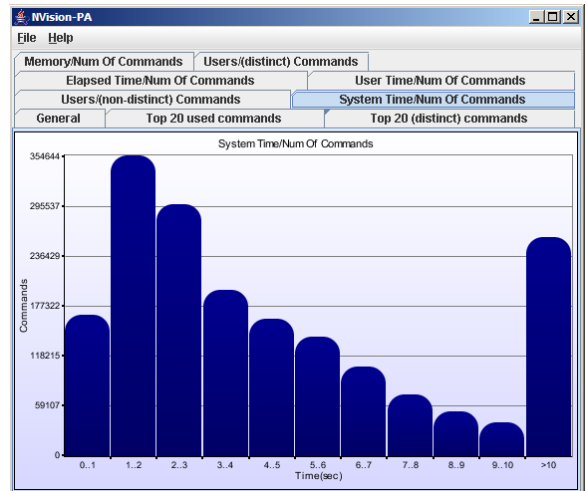
**(B)**

**Figure 10.** NVision-PA Distribution of <u>System</u> Time Over Commands: (A) Internet PA Log File versus (B) HPC Cluster PA Log File.

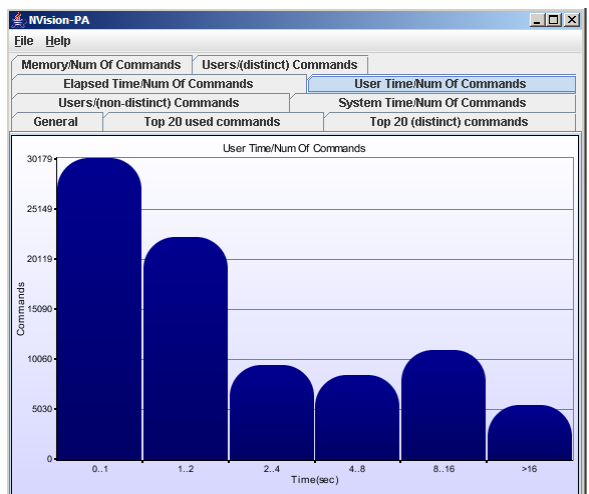
**(A)**

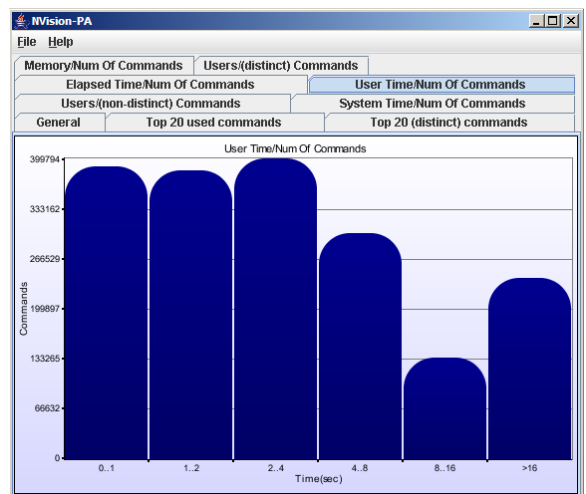
**(B)**

**Figure 11.** NVision-PA Distribution of <u>User</u> Time Over Commands: (A) Internet PA Log File versus (B) HPC Cluster PA Log File

commands. The distribution of system time for the Internet server commands decays quickly from this peak while the distribution of system time for HPC cluster commands decays gradually to another bimodal peak. The second peak for both system time distributions is commands that execute with a system time over 10 seconds. Roughly 13% of HPC cluster commands and 8% of Internet server commands execute with a system time over 10 seconds.

Figure 11 presents the NVision-PA "User Time/Number of Commands" tab results that refer to the distribution of user time over all commands[7] in the selected input log. The desired insight from this tab report is the distribution of user time per command for different environments during the period contained in the selected input log. In other words, from this report we get an idea of how long a user is engaged with a particular process. For example, if users run commands like *ls* or *cp* then the average user time will be low while if users execute new shells or tools like *gdb* or *emacs* then the distribution of user times will be skewed higher. We can see from Figure 11A that the user time of Internet server commands peaks with 35% of all commands falling within the range 0-1 seconds with an exponential drop-off from this peak. In Figure 11B we see that the user time of HPC cluster commands does not peak but rather more closely resembles a uniform distribution with these percentages of commands: 21% (0-1sec), 21% (1-2sec), 22% (2-4sec), 16% (4-8sec), 7% (8-16sec), and 13% (>16sec).

Figure 12 presents the NVision-PA "Elapsed Time/Number of Commands" tab results

---

[7] Each command execution instance is counted, the same command executed *x* times is counted *x* times.

that refer to the distribution of elapsed time over all commands[8] in the selected input log. The desired insight from this tab report is the distribution of elapsed time per command for different environments during the period contained in the selected input log. This report merges the two previous time reports (Figures 10 and 11) since it always holds that elapsed time is always great than the sum of system time plus user time. Elapsed time also contains delays caused by the scheduling of processes which may or may not be significant in different environments. In Figure 12A the elapsed time distribution for commands executed on the Internet server shows 4 peak modes: 53% (0-2, 2-4, 4-6 seconds), 6% (10-20sec), 8% (100-200sec), and 16% (> 400sec) – which when combined accounts for 83% of all commands. In Figure 12B, the elapsed time distribution for commands executed on the HPC cluster shows 4 similar peak modes: 25% (2-4, 4-6 seconds), 20% (10-20sec), 6% (100-200sec), and 8% (> 400sec) – which when combined accounts for 59% of all commands.

Figure 13 presents the NVision-PA "Memory/Number of Commands" tab results that refer to the distribution of memory usage over all commands[9] in the selected input log. The desired insight from this tab report is the distribution of memory usage per command for different environments during the period contained in the selected input log. In other words, this tab gives us an idea of the memory demands of users executing jobs in different environments. The metric in processing accounting for measuring memory is *ac_mem,* the average amount of memory in units of 8K (pages) that is used by the command process. Before comparing data for the different environments, it should be noted that the $x$ axis in Figures 13A and 13B have different labels. In Figure 13A, the memory usage distribution for commands executed on the Internet server shows 2 peak modes: 53% (100-500 memory pages) and 32% (> 1000 memory pages) – which when combined accounts for 85% of all commands.    In Figure 13B, the memory usage distribution for commands executed on the HPC cluster shows only 1 peak mode which characterizes most of the commands: 94% (2000-7000 memory pages).    Comparing Figures 13A and 13B reveals that HPC cluster users execute commands with distinctly more memory requirements than commands executed by users in an Internet server environment.

---

[8] Each command execution instance is counted, the same command executed *x* times is counted *x* times.
[9] Each command execution instance is counted, the same command executed *x* times is counted *x* times.

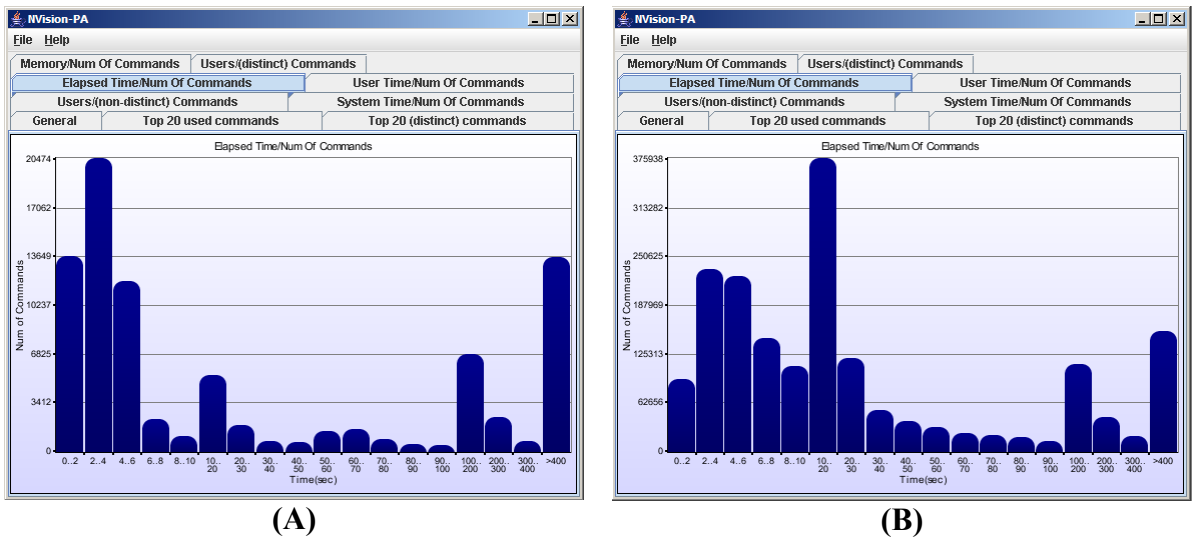

**Figure 12.** NVision-PA Distribution of <u>Elapsed</u> Time Over Commands: (A) Internet PA Log File versus (B) HPC Cluster PA Log File

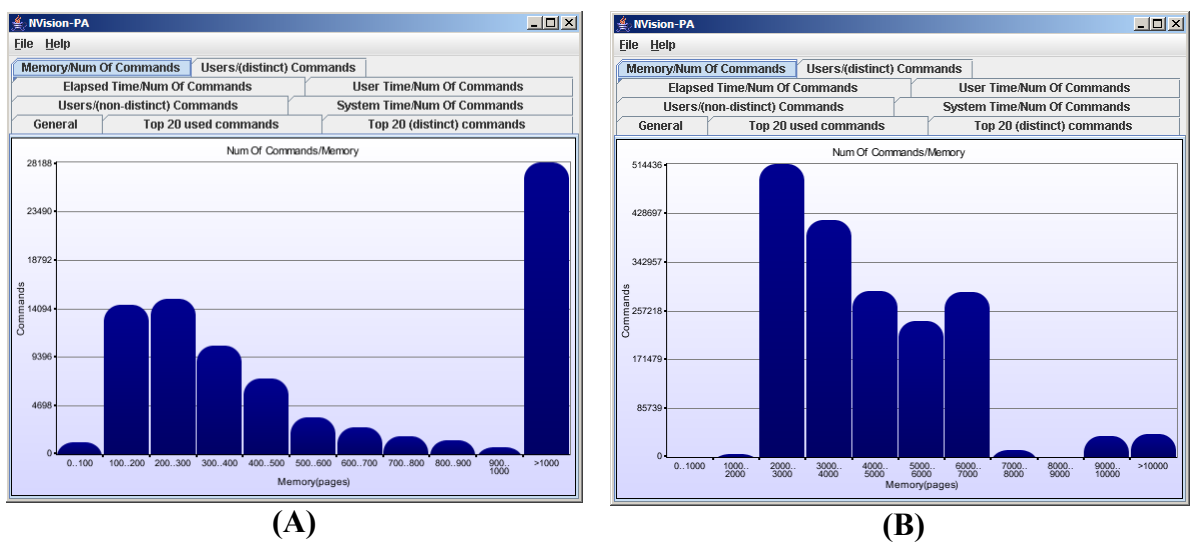

**Figure 13.** NVision-PA Distribution of Average Memory Usage Over Commands: (A) Internet PA Log File versus (B) HPC Cluster PA Log File  *(Note x-axis differs in 13A and 13B)*

Lastly, this is the second data set of Internet server/HPC cluster process accounting logs we have analyzed, the first data set was briefly described in [9].  The results from both data sets are consistent so we believe them to be accurate characterizations of their environments.    Whether the environments from which we obtained the process accounting logs is typical or atypical of general Internet server or general HPC cluster environments is an open question.   A benchmark of standard Internet server or HPC cluster process accounting does not exist and may not be feasible given the wide variety of legitimate command behaviors.    However, what we have presented here is feature information that can be fed to a pattern classification algorithm, such as SVM, in order to enhance discrimination of command behaviors for the particular environment measured. This same procedure using NVision-PA may be used to enhance discrimination of command behaviors for any set of environments.

## 5.0 Summary

NVision-PA makes available sophisticated analysis of process accounting logs using a Java GUI portable to most environments. Specifically in this paper we report results from NVision-PA for comparing two process accounting data sets; one from an Internet server environment and one from an HPC cluster environment. The features revealed by NVision-PA in these data sets are an incremental but significant next step in developing a real-time masquerade detector for the HPC cluster environment based on command behavior. Beyond masquerade detection, the analysis capability of NVision-PA also promises to be useful in tuning complex system environments for workloads that can be characterized by command behaviors.

## References

[1]   A. Cockcroft, *Processing Accounting Data into Workloads*, Sun BluePrints, 1999.

[2]   K. Gilbertson, "Process Accounting," *Linux Journal*, December 12, 2001.

[3]   V. Hazelwood, "Unix Accounting Magic," S*ysAdmin Magazine*, March 1999.

[4]   *IEEE International Symposium on Workload Characterization – IISWC (formerly known as the Workshop on Workload Characterization –WWC),* 2005 IISWC website: <http://www.iiswc.org/iiswc2005/home.html>

[5]   K. Luo, Y. Li, C. Ermopoulos, W. Yurcik, and A. Slagell "Scrub-PA: A Multi-Level Multi-Dimensional Anonymization Tool for Process Accounting," *ACM Computing Research Repository (CoRR) Technical Report cs.CR/0601079* , January 2006.

[6]   D.A. Menasce, "Workload Characterization," *IEEE Internet Computing*, Sept/Oct. 2003.

[7]   S. Peisert, "Forensics for System Administrators," *Usenix ;Login*, August 2005.

[8]   A. Tam. *Enabling Process Accounting on Linux HOWTO, version 1.1*, 2001. <http://www.tldp.org/HOWTO/ProcessAccounting/>

[9]   W. Yurcik and C. Liu, "A First Step Toward Detecting SSH Identity Theft in HPC Cluster Environments: Discriminating Cluster Masqueraders Based on Command Behavior," *1st International Workshop on Cluster Security (Cluster-Sec)* held in conjunction with the *5th IEEE Intl. Symposium on Cluster Computing and the Grid (CCGrid)*, Cardiff U.K., 2005.